# Chiral anomaly, triangle loop and the $\gamma\gamma^* \to \pi^0$ form factor

Phạm Trí Năng[a] and Phạm Xuân Yêm[b]

[a]Centre de Physique Théorique, CNRS, Ecole Polytechnique, 91128 Palaiseau, France

[b] Laboratoire de Physique Théorique et Hautes Energies, Université Pierre &Marie Curie, Unité associée au CNRS, 75005 Paris, France

**Abstract**

The recent BaBar measurements of the $\gamma + \gamma^* \to \pi^0$ form factor show spectacular deviation from perturbative QCD computations for large space-like $Q^2$. At 34 GeV$^2$ the data are more than 50% larger than theoretical predictions.
Stimulated by these new experimental results, we revisit our previous paper on triangle loop effects related to chiral anomaly, and apply our method to the $\gamma + \gamma^* \to \pi^0$ form factor measured in the single tag mode $e^+ + e^- \to e^+ + e^- + \pi^0$ with one highly virtual photon $\gamma^*$. The resultant form factor $F(Q^2)$ – which depends on only one parameter (the mass $m$ of up, down quark circulating in the triangle loop) behaves like $(m^2/Q^2) \times [\ln(Q^2/m^2)]^2$ – shows a striking agreement with BaBar data for $m \approx 132$ MeV. The rising logarithm squared form factor, surprisingly unnoticed in the literature, is in sharp contrast with the rather flat ones derived from perturbative QCD approaches.

**Introduction**

As is well known, the Adler- Bell- Jackiw (ABJ) anomaly [1] is at the heart of the $\pi^0 \to \gamma + \gamma$ decay. It begins with a paradox [2] found by Sutherland and Veltman who showed that application of Current Algebra and PCAC (Partial Conservation of the Axial Current) [1] gave a vanishing rate to this decay, in contradiction with experiments. By solving the paradox, ABJ discovered chiral anomaly and opened a new window for the understanding of subtle quantum effects. It provides a beautiful manifestation of the ultraviolet-infrared correlation in gauge field theories: short distance singularities manifest themselves in low energy theorems [3] which can give rise to spectacular physical consequences experimentally observed.
What should be changed when one of the photon $\gamma^*$ is off the mass shell, does the ABJ anomaly still play its crucial role with $\gamma^*$?
The virtual $\gamma^*$ appears either in its time-like form (processes $e^+ + e^- \to \gamma^* \to \pi^0 + \gamma$ and its straightforward generalisation to the weak boson decays $Z^0 \to \pi^0 + \gamma$, $W^\pm \to \pi^\pm + \gamma$) or in its space-like one (the recent BaBar single tag mode [4] $e^+ + e^- \to e^+ + e^- + \pi^0$ in which one of the outgoing leptons emitting a highly virtual photon $\gamma^*$ is detected, the other untagged lepton is scattered at very small angle, its momentum transfer is nearly zero corresponding to a quasi real photon $\gamma$).

We argue that when the two photons interact with $\pi^0$, no matter they are on or off the mass shell, it is very natural to treat them on the same footing. The reason is that the pion, a Nambu-Goldstone boson of the spontaneous chiral symmetry breaking in QCD, is intimately related to PCAC which is **altered** by the ABJ anomaly issued from ultraviolet divergence in triangle loops. Since the two photons are simply external particles outside the loops, their nature (real or virtual) seems to be dynamically unconcerned by loop integrals.

For the comprehension of the problem, let us briefly recall some conceptually important points:

### 1- PCAC

For massless fermions (i.e. up and down quarks), both Quantum Chromodynamics (QCD) and Quantum Electro-Weak interactions are governed by chiral symmetry associated with the separate number conservation of left-handed and right-handed fermions. The corresponding Noether currents (and their linear combinations, i.e. the vector current and the axial-vector current) are both conserved. Note that Dirac equation ensures that the former is always conserved whether fermions are massless or not, the latter is conserved only for massless fermions. The chiral symmetry is generated by the axial vector current $A^\mu(x)$ with $\partial_\mu A^\mu(x) \sim 2m$ which tends to 0 as the fermion mass $m \to 0$. Since in QCD, quarks and antiquarks have strong attractive interactions at low energy and if they are massless, then the energy cost creating an quark-antiquark pair is small, thus we expert that the vacuum of QCD contains a condensate of quark-antiquark pairs, characterized by a non zero vacuum expectation value which is a manifestation of spontaneous chiral symmetry breaking. The Goldstone theorem tells us that we would find three spin 0 massless particles associated to three components of the isospin SU(2) axial vector currents. The real strongly interacting particles do not contain any massless spin 0 mesons, however among some hundreds of hadrons we do observe an isospin triplet of extremely light mesons, the pions. They are the quasi Goldstone bosons with mass $m_\pi \approx 0$. Since pion is a pseudoscalar meson, they can be created by the axial vector current: $\langle 0| A^\mu(x) |\pi(q)\rangle = -i q^\mu f_\pi e^{-i q.x}$, where $f_\pi \approx 93$ MeV is the pion decay constant associated with the neutral $\pi^0$. For charged $\pi^\pm$ pions, their decay constant is $\sqrt{2} f_\pi$ usually denoted as $F_\pi \approx 132$ MeV measured from $\pi^\pm \to \mu^\pm + \nu$.

Thus $\langle 0| \partial_\mu A^\mu(x) |\pi(q)\rangle = m_\pi^2 f_\pi e^{-i q.x}$ and PCAC may be formulated in another equivalent way as an operator equation:

$$\partial_\mu A^\mu(x) = m_\pi^2 f_\pi \phi(x) \to 0 \quad \text{as} \quad m_\pi \to 0 \qquad (1)$$

Here $\phi(x)$ is the pion field normalized as $\langle 0| \phi(x)|\pi(q)\rangle = e^{-i q.x}$. The axial vector current is conserved in the limit of massless pion, that is the physical meaning of PCAC (Partial Conservation of the Axial Current).

### 2- Triangle loop and Anomaly

However, in general the true picture is more complicated because of quantum effects due to higher order radiative corrections. Anomaly is a manifestation of quantum effects, it tells us that triangle loop destroys the chiral symmetry of the classical equations of motion. That in turn can violate gauge invariance or equivalently cannot respect the axial Ward identity.

PCAC given by Eq.(1) is actually spoiled by loop corrections and becomes incompatible with gauge invariance or conservation of currents. This happens because triangle diagrams supply a nonzero term on the right hand side of Eq. (1), called ABJ anomaly :

$$\partial_\mu A^\mu (x) = m_\pi^2 f_\pi \phi(x) + S\, e^2/(16\pi^2)\, \varepsilon_{\alpha\beta\gamma\delta} F^{\alpha\beta} F^{\gamma\delta} \qquad (2)$$

We recognize in Eq.(2) the familiar $1/16\pi^2$ of one-loop integrations as well as the higher order $e^2$ expansion. Here $F^{\mu\nu} = \partial^\mu J^\nu - \partial^\nu J^\mu$ is the electromagnetic field strength tensor. The coefficient S depends on the theoretical model of fermions from which the axial vector current $A^\mu$ as well as the electromagnetic vector current $j^\mu$ are built and coupled to [5].
Since anomalies destroy gauge invariance, their total absence is a sine que non for gauge theories to be renormalizable. Anomalies coming from separate sectors of quark and lepton must combine in such a way that they cancel each other. This constitutes an important constraint on any renormalizable theory. The anomaly-free property [6] indeed occurs in the Glashow-Salam-Weinberg electroweak gauge theory due to the equal numbers of quark and lepton doublets.

### 3- Anomaly effects

Although the overall anomalies must be mutually cancelled in renormalizable gauge field theories, nevertheless a particular anomaly term alone somehow could have remarkable physical consequences, the famous one being the $\pi^0 \to \gamma + \gamma$ decay for which the three colours $N_c$ of quarks exhibit their glaring evidence [5] by giving the correct number 1/2 to the coefficient S in Eq.(2).
The relevance of chiral anomaly in the processes involving a pion and two photons (no matter they are on or off mass shell) relies on our identification of pions as the Nambu-Goldstone bosons of spontaneous chiral symmetry breaking of QCD in which PCAC plays a central role. We apply the method used in our previous works [7] on the cancellation of chiral anomaly effects in $Z^0 \to \pi^0 + \gamma$, $W^\pm \to \pi^\pm + \gamma$, $e^+ + e^- \to \gamma^* \to \pi^0 + \gamma$ to the new one $\gamma + \gamma^* \to \pi^0$ corresponding to the recent BaBar single tag mode $e^+ + e^- \to e^+ + e^- + \pi^0$.
The $\gamma\gamma^* \to \pi^0$ amplitude $\langle \pi^0(q) |T| \gamma(k), \gamma^*(k')\rangle$ has the general form $\varepsilon_\mu(k)\varepsilon_\nu(k')N^{\mu\nu}(k, k')$ where $N^{\mu\nu}(k', k) = e^2 F(k', k) Y^{\mu\nu}$ with the kinematic factor $Y^{\mu\nu} = \varepsilon^{\mu\nu\alpha\beta} k_\alpha k'_\beta$, $e$ is the electric charge and $F(k', k) \equiv F(Q^2)$ is the $\gamma\gamma^* \to \pi^0$ transition form factor we consider in this paper.
The $\gamma\gamma^* \to \pi^0$ amplitude for off mass shell pion, as given by the Lehmann-Symanzik-Zimmermann procedure, can be represented by

$$N^{\mu\nu}(k', k) = (m_\pi^2 - q^2) \int d^4x\, d^4y\, exp(i\,kx + i\,k'y)\,\langle 0 | T(J^\mu(x) J^\nu(y) \phi(0)) | 0\rangle \qquad (3)$$

where $q = k + k'$ is the pion momentum.

Using Eq.(2) which relates the divergence of the axial vector current $\partial_\tau A^\tau(x)$ to the pion field $\phi(x)$, naturally arises a three index pseudotensor $R^{\tau\mu\nu}(k', k)$ defined by:

$$R^{\tau\mu\nu}(k', k) = i \int d^4x\, d^4y\, exp(i\,kx + i\,k'y)\,\langle 0 | T(J^\mu(x) J^\nu(y) A^\tau(0)) | 0\rangle \qquad (4)$$

Combining Eqs. (2), (3), (4) and using the standard Current Algebra technique, we obtain the off shell pion amplitude [7] :

$$N^{\mu\nu}(k', k) = (m_\pi^2 - q^2)/(f_\pi m_\pi^2)\, [q_\tau R^{\tau\mu\nu}(k', k) - S e^2/(2\pi^2)\, Y^{\mu\nu}] \qquad (5)$$

Eq. (5) represents the starting point of our analysis and the problem is concentrated on how to compute $q_\tau R^{\tau\mu\nu}$ (k', k).

The remarkable kinematic properties of $R^{\tau\mu\nu}$ (k', k) having odd parity, satisfying Bose symmetry and transverse to k and k' ($k'_\mu R^{\tau\mu\nu}$ (k', k) = 0 = $k_\nu R^{\tau\mu\nu}$ (k', k)) tell us that its divergence $q_\tau R^{\tau\mu\nu}$ (k', k) is proportional to $k^2$ and $k'^2$. Therefore $q_\tau R^{\tau\mu\nu}$ (k', k) must be non zero [1, 7, 8] provided that one of the photon is virtual, a model-independent result actually. Only when both photons are real that $q_\tau R^{\tau\mu\nu}$ (k', k) vanishes, thus showing the Sutherland-Veltman paradox since the "*anomalous*" term $Se^2/(2\pi^2) Y^{\mu\nu}$ of Eq.(5) was absent. This vanishing property of $q_\tau R^{\tau\mu\nu}$ (k', k) will be confirmed below in the particular case of triangle loop computation.

Since the pion is the only low mass particle, we should distinguish at low four-momentum q, the one pion-pole contribution to $R^{\tau\mu\nu}$ (k', k) from the remainder corresponding to a direct coupling to quarks of the three currents $J^\mu$, $J^\nu$ and $A^\tau$ denoted by $D^{\tau\mu\nu}$(k', k). This decomposition is reminiscent of the derivation of the Goldberger-Treiman (GT) relation [9] and also used by Bell and Jackiw in their treatment of the $\pi^0 \to \gamma + \gamma$ decay [1]:

$$R^{\tau\mu\nu} (k', k) = D^{\tau\mu\nu} (k', k) - f_\pi /(q^2 - m^2_\pi) [q^\tau N^{\mu\nu}(k',k)] \qquad (6)$$

Eq. (5) together with Eq. (6) yields:

$$N^{\mu\nu} (k', k) = (1/f_\pi)\{q_\tau D^{\tau\mu\nu} (k', k) - Se^2/(2\pi^2) Y^{\mu\nu}\} \qquad (7)$$

The problem then reduces to the computation of $q_\tau D^{\tau\mu\nu}$ (k', k).

For high-energy virtual photon scattering, it is plausible to apply the quark-parton model to $D^{\tau\mu\nu}$ (k', k) and naturally arise at one-loop level the triangle diagrams in which the three currents $J^\mu$, $J^\nu$, $A^\tau$ directly couple to the up and down quarks. We assume from now on that $D^{\tau\mu\nu}$ (k', k) is given by contributions from triangle diagrams:

$$D^{\tau\mu\nu} (k', k) = \sum_{j=u, d\ quarks} S_j T_j^{\tau\mu\nu} (k', k) \equiv Se^2 T^{\tau\mu\nu} (k', k) \qquad (8)$$

In the sum over up and down quarks, all the couplings to the three currents are explicit as follows: S = $S_u + S_d$ = $N_c (g_u e_u^2 + g_d e_d^2)$ where $g_u = - g_d = ½$ are the couplings of the up and down quarks to the axial vector current $A^{\tau=3}$, and $e_u$ = 2/3, $e_d$ = −1/3 (in units of the proton charge +e) are charges of u, d quarks coupled to the electromagnetic currents $J^\mu$, $J^\nu$, and $N_c$ = 3 is the number of colors of quark. So S = 1/2. Without $N_c$, one would get in quark model a $\pi^0 \to \gamma + \gamma$ decay rate smaller by a factor of 9 compared to experiments.

### 4- A brief calculation

General features of the triangle loop calculations are already outlined in [1,8], nevertheless the resultant Eqs. (13) and (14) issued from exact calculations shown below may be still overlooked [7]. It has two parts, the finite P(k', k) part and the divergent one related to anomaly:

$$q_\tau T^{\tau\mu\nu} (k', k) = \{P(k', k) + 1/(2\pi^2)\} Y^{\mu\nu} \qquad (9)$$

where $\quad P(k', k) = (m^2/\pi^2) \int_0^1 dx \int_0^{1-x} dy \; 1/D(k', k, x, y)$ $\quad\quad\quad$ (10)

The denominator $D(k', k, x, y) = k^2 y(1-y) + k'^2 x(1-x) + 2 k.k' \, x y - m^2$ comes from the Feynman parameterization [9] of the product $1/[(p-k)^2 - m^2] \times [p^2 - m^2] \times [(p+k')^2 - m^2]$ in the denominators of the 3 quark propagators and integrated over $d^4p$ in the loop.

The first term $P(k', k)$ in Eq. (9) comes from the p-independent term in the numerator of the product of these 3 propagators. This p-independent term proportional to $m^2$ yields a finite contribution since it is of the form $\sim d^4p/p^6$. The p-dependent term in the numerator of the product of these 3 quark propagators yields a logarithmic divergence $\sim d^4p/p^4$ in the integral. The regularization of this logarithmic divergence yields the second term $1/(2\pi^2)$ which is precisely the ABJ anomaly [10]. The typical kinematic factor $Y^{\mu\nu}$ comes from the trace of the product of four and six Dirac matrices with $\gamma_5$.

The easiest way to calculate anomaly is the use of the Pauli–Villars regularization to deal with divergences, since dimensional regularization [9] has some ambiguities in the extension of $\gamma_5$ to $n$ dimensions. The Pauli–Villars regularization consists of considering $q_\tau T^{\tau\mu\nu}(k', k)$ to be a function of the fermion mass circulating in the loop, let us write it as $q_\tau T^{\tau\mu\nu}(k', k, m)$. The regularized amplitude – defined as the difference between $q_\tau T^{\tau\mu\nu}(k', k, m)$ and the same amplitude $q_\tau T^{\tau\mu\nu}(k', k, M)$ taken at some big mass $M$ – is finite since the divergence of the former is cancelled by that of the latter.

$$q_\tau T^{\tau\mu\nu}(k', k)_{regular} = q_\tau T^{\tau\mu\nu}(k', k, m) - q_\tau T^{\tau\mu\nu}(k', k, M).$$

At the end, one lets $M \to \infty$ in the finite part of $q_\tau T^{\tau\mu\nu}(k', k, M)$, which is nothing else than the ABJ anomaly. Since the finite part of $q_\tau T^{\tau\mu\nu}(k', k, m)$ is $P(k', k)Y^{\mu\nu}$, the finite part of $q_\tau T^{\tau\mu\nu}(k', k, M)$ is simply obtained by the substitution $m \leftrightarrow M$ in Eq. (10), i.e.

$$(M^2/\pi^2) \int_0^1 dx \int_0^{1-x} dy \; 1/[k^2 y(1-y) + k'^2 x(1-x) + 2 k.k' \, x y - M^2] \quad\quad (11)$$

then let $M \to \infty$ in Eq. (11) which yields $-(1/2\pi^2)$. Thus in this limit $M \to \infty$, the quantity $q_\tau T^{\tau\mu\nu}(k', k)_{regular}$ as given by the r.h.s. of Eq. (9) is $\{P(k', k) + 1/(2\pi^2)\} Y^{\mu\nu}$.

It is interesting to check that when both photons are real: $k^2 = k'^2 = 0$, $2 k.k' = m_\pi^2 = O(q^2)$, Eq.(10) yields $P(k', k) = -(1/2\pi^2) + O(q^2)$. Then from Eq. (9), $q_\tau T^{\tau\mu\nu}(k', k) = O(q^2)$ and we recover the Sutherland-Veltman paradox. The second term $Se^2/(2\pi^2)$ in Eq.(7) which is the ABJ anomaly then plays its central role by giving the correct rate to $\pi^0 \to \gamma + \gamma$ decay.

## 5- The form factors

The three processes $\gamma\gamma^*\pi^0$ respectively with real, time-like and space-like virtual photon are described on the same footing by the familiar triangle diagrams.

Since $N^{\mu\nu}(k', k) = e^2 F(k', k) Y^{\mu\nu}$, then using Eqs. (7), (8) and (9), the corresponding form factors $F(k', k)$ in these 3 cases are given by:

$$F(k', k) = (1/4\pi^2)(1/f_\pi) \int_0^1 dx \int_0^{1-x} dy \; 2m^2/D(k', k, x, y)$$

Here $m$ is the up, down quark mass taken to be the same, with the denominator $D(k', k, x, y) = k^2 y (1-y) + k'^2 x (1-x) \pm 2\, k.k'\, x y - m^2$. The $\pm$ factor in $2\, k.k' xy$ of the denominator $D(k', k, x, y)$ reflects respectively the space (time) like virtual photon of momentum k': $-k'^2 = Q^2 > 0$ (for space-like $\gamma^*$) and $k'^2 = s \geq 0$ (for time-like and light-like $\gamma^*$). Also respectively $k' \pm k = q$, the pion momentum. We take $q^2 = m^2_\pi = 0$ so that $2\, k.k' = s$ (or $Q^2$) in both cases of time (or space) like photon.

Integrating first over the y variable, we obtain

$$F(k', k) = (1/4\pi^2)(1/f_\pi)(2m^2/|k'^2|) \int_0^1 \ln[1 - (k'^2/m^2) x(1-x)]\, dx/x$$

When the $x$ integration is done, the transition form factors of these processes are found to be:

a- **Real photons** $(k'^2 = 0, k^2 = 0)$

$$F(0,0) = -(1/4\pi^2)(1/f_\pi) \qquad (12)$$

from which the rate $\Gamma(\pi^0 \to \gamma + \gamma) = [e^2 F(0,0)]^2\, m^3_\pi/64\pi = \alpha^2/(64\pi^3)(m^3_\pi/f_\pi^2)$

is in excellent agreement with experiments.

b – **Time-like photon** $(k'^2 = s > 0, k^2 = 0)$ :

$$F(s, 0) = +(1/4\pi^2)(1/f_\pi)(2m^2/s)[Sp(-2/(1-\rho)) + Sp(-2/(1+\rho))]$$

with $\rho = \sqrt{[1 - (4m^2/s)]}$ and Sp $(\xi)$ is the Spence (or dilogarithm) function [11] behaving like a product of two logarithms. It can be shown that

$$2[Sp(-2/(1-\rho)) + Sp(-2/(1+\rho))] = [\ln(1-\rho)/(1+\rho) - i\pi]^2$$

The presence of the imaginary part $2 i\pi \ln(1-\rho)/(1+\rho)$ is due to the fact that the argument $1 - (s/m^2) x(1-x)$ of $\ln[1 - (s/m^2) x(1-x)]$ can be negative, physically it means that the time-like $\gamma^*$ can decay into a real quark-antiquark pair which then converts into $\pi^0 + \gamma$.

For $s \gg 4m^2$, $[\ln(1-\rho)/(1+\rho) - i\pi]^2 \to [\ln(s/m^2) - i\pi]^2$

$$F(s, 0) = (1/4\pi^2)(1/f_\pi)(m^2/s)[\ln(1-\rho)/(1+\rho) - i\pi]^2$$

$$\to (1/4\pi^2)(1/f_\pi)(m^2/s)[\ln(s/m^2) - i\pi]^2 \qquad (13)$$

c- **Space-like photon** $(k'^2 = -Q^2 \leq 0, k^2 = 0)$ :

$$F(Q^2, 0) = +(1/4\pi^2)(1/f_\pi)(2m^2/Q^2)[Sp(2/(\rho'-1)) + Sp(-2/(\rho'+1))]$$

with $\rho' = \sqrt{[1 + (4m^2/Q^2)]}$. Also it turns out that

$2[Sp(2/(\rho'-1)) + Sp(-2/(\rho'+1))] = [\ln(\rho'-1)/(\rho'+1)]^2 \to [\ln(Q^2/m^2)]^2$ for $Q^2 \gg 4m^2$, thus

$$Q^2F(Q^2,0) = (1/4\pi^2)(m^2/f_\pi)[\ln(\rho'-1)/(\rho'+1)]^2 \to (1/4\pi^2)(m^2/f_\pi)[\ln(Q^2/m^2)]^2 \quad (14)$$

However for small $Q^2$, $Q^2F(Q^2, 0) \to 0$ as $Q^2 \to 0$ due to Eq. (12).

The curve $Q^2F(Q^2, 0) = (1/4\pi^2)(m^2/f_\pi) [\ln(Q^2/m^2)]^2$ is plotted in Fig. 1 with m = 132 MeV for all values of $Q^2$ ($0 \leq Q^2 < \infty$) and compared to the recent BaBar measurement of the $\gamma \gamma^* \to \pi^0$ form factor. Even for small $Q^2$, the curve fits well the CLEO data.

**Conclusion**. Based on ABJ chiral anomaly and triangle loop, the striking feature of our calculation is the logarithm squared $[\ln(Q^2/m^2)]^2$ of the form factors which is able to explain the large rise for the quantity $Q^2F(Q^2)$ of BaBar data. This behaviour seems to be unnoticed in the literature. Another approaches using perturbative QCD [12] lead to a rather flat shape since the pion distribution amplitude $\phi(x, Q^2)$ used in these methods does not have the typical logarithm squared issued from our exact calculation of triangle loop integrals.

A.E.Dorokhov, Phys. Part. Nucl. Lett. **7,** 229 (2010), arXiv:hep-ph/0905.4577 and arXiv:hep-ph/1003.4693].

[8] L. Rosenberg, Phys. Rev. **129**, 2786 (1963)

[9] See text books, for instance Quang Ho-Kim and Pham Xuân Yêm, Elementary Particles and their Interactions, Concepts and Phenomena, Springer, Berlin, NewYork (1998): page 437 for the GT relation, page 478 for the dimensional regularization and pages 651-652 for the Feynman parameterization and loop integrations.

[10] By shifting the integration variable p → p ± k (or p → p ± k'), the divergence in triangle loop happens to vanish identically, so at first sight the shift seems to be a miraculous trick to handle the divergences. However the fallacy in this illegal exercise for divergent integrals is that it changes the integrals by a finite amount. There always exists a finite anomaly term independently of the regularization procedures.
See text books, for instance P. Frampton, Gauge Field Theories, The Benjamin/Cummings Pub. Com. (1987), pages 202-205.

[11] $Sp(x) = \int_0^x du \ln(1+u)/u$, $Sp(-x) = \int_0^x du \ln(1-u)/u$, see text books, for instance S. Pokorski, Gauge Field Theories, Cambridge Monographs on Mathematical Physics (2000), page 562.

[12] G.P. Lepage and S.J. Brodsky, Phys. Rev. D **22**, 2157 (1980); P. Kroll and M. Raulfs, Phys. Lett.B **387**,848 (1996); A.P. Bakulev, S.V. Mikhailov, and N.G. Stefanis, Phys. Rev.D **67,** 074012 (2003), Phys. Lett.B **578**, 91 (2004).
For recent developments, see S .V. Mikhailov, and N.G. Stefanis, Nucl.Phys. B **821**, 291 (2009). P.Kroll [arXiv:hep-ph/1012.3542]; S. S. Agaev, V. M. Braun, N. Offen and F.A. Porkert [arXiv:hep-ph/1012.4671].

**Figure caption:** Predictions from Triangle loop (solid line) and Asymptotic Perturbative QCD (horizontal dash line) compared with the BaBar and CLEO measured values for Q^2*F(Q^2).

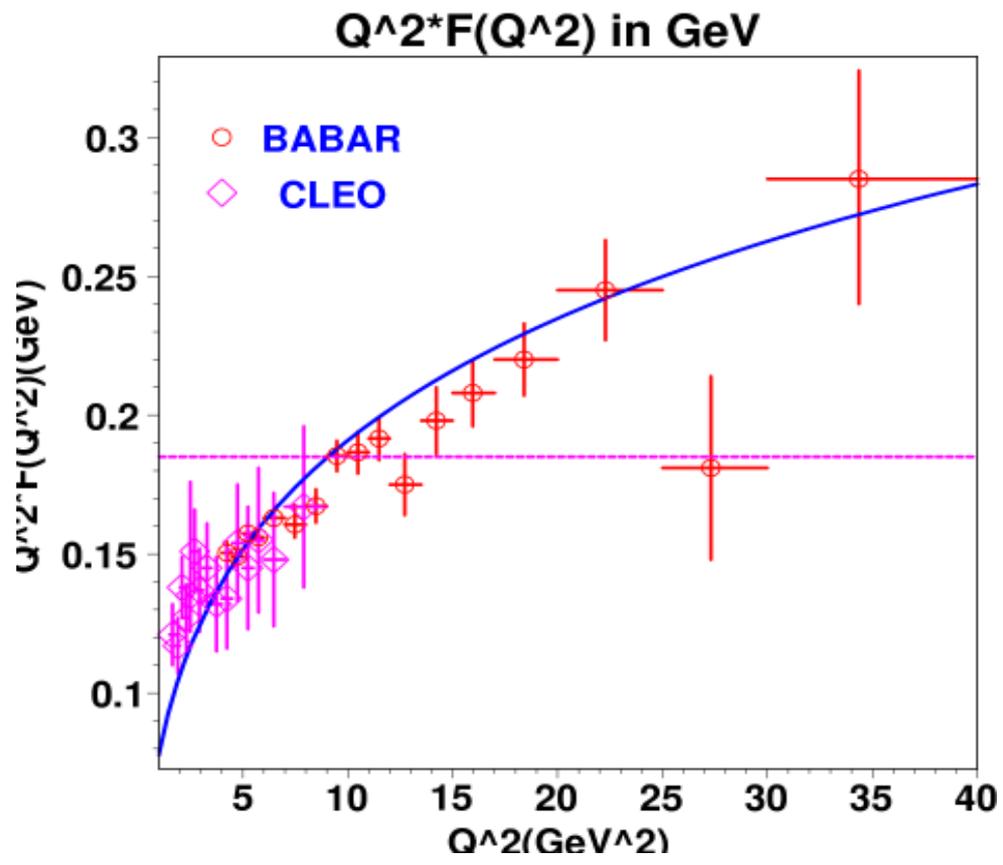

Fig. 1